\documentclass[12pt]{article}
\usepackage{psfrag,epsfig,amsmath,amssymb}

\makeatletter
\renewcommand{\section}{\setcounter{equation}{0}\@startsection
  {section}%
  {1}%
  {0pt}%
  {-1\baselineskip}%
  {0.4\baselineskip}%
  {\bfseries\large}}%
\renewcommand{\subsection}{\@startsection
  {subsection}%
  {2}%
  {0pt}%
  {-0.75\baselineskip}%
  {0.2\baselineskip}%
  {\bfseries}}%
\renewcommand{\subsubsection}{\@startsection
  {subsubsection}%
  {3}%
  {0pt}%
  {-0.5\baselineskip}%
  {0.1\baselineskip}%
  {\sc}}%
\makeatother

\makeatletter
 \newcommand\figcaption{\def\@captype{figure}\caption}
\makeatother

\def\a{\alpha}
\def\b{\beta}

\def\m{\mu}
\def\n{\nu}

\def\s{\sigma}

\def\th{\theta}

\def\Dirac{{D\mkern-12mu/}}

\def\prslash{{\partial\mkern-9mu/}}

\def\prslash{{\partial\mkern-9mu/}}    

\def\bp{\text{\tiny{BPST}}}
\def\id{{\rm{I}\!\rm{I}}}

\def\id3x{\int\!\! d^3\!\vec{x}}
\def\idx{\int\!\! d^4\!x}

\def\rig>{\right>}

\newcommand{\bea}{\begin{eqnarray}}
\newcommand{\eea}{\end{eqnarray}}
\newcommand{\beann}{\begin{eqnarray*}}
\newcommand{\eeann}{\end{eqnarray*}}
\newcommand{\ba}{\begin{array}}
\newcommand{\ea}{\end{array}}

\newcommand{\Tr}{\mathbf{Tr}}

\def\Psib{\bar{\Psi}}
\def\g5{\gamma_{5}}

\def\prslash {{\partial\mkern-9mu/}}  

\def\idx3{\int\! d^{3}\!\vec{x}\,}
\def\idx{\int\! d^{4}\!x\,}

 \def\Psib{\bar{\Psi}}

 \def\Dirac{{D\mkern-12mu/}\,}

 \def\prslash {{\partial\mkern-9mu/}}  

\def\bk {\bar{k}} 
 \def\bp {\bar{p}}
 \def\bq {\bar{q}} 
 \def\bg {\bar{\gamma}}
 \def\bdelta {\bar{\delta}}

 \def\Db {{\partial}_{\beta}}

 \def\ab {a_{\beta}}
 \def\am {a_{\mu}}

 \def\g {\gamma}

 \def\a {\alpha}
\def\b {\beta}

 \def\s {\sigma}

 \def\Tr{\text{Tr}}

\begin{document}
\begin{titlepage}
\hfill{NSF-KITP-09-113}\\
\rightline{FTI/UCM 91-2009}\vglue 10pt
\begin{center}

{\Large \bf Noncommutative GUT inspired theories and the UV finiteness of the fermionic four point functions}\\
\vskip 1 true cm {\rm C.P. Mart\'{\i}n$^{\dagger,}$\footnote{E-mail: carmelo@elbereth.fis.ucm.es}
and  C. Tamarit$^{\dagger\dagger,}$}\footnote{E-mail: tamarit@kitp.ucsb.edu}
\vskip 0.3 true cm $^\dagger${\it Departamento de F\'{\i}sica Te\'orica I,
Facultad de Ciencias F\'{\i}sicas\\
Universidad Complutense de Madrid,
 28040 Madrid, Spain}
 \vskip 0.3 true cm $^{\dagger\dagger}${\it Kavli Institute for Theoretical Physics, University of California\\
 Santa Barbara, CA, 93106-4030, USA}\\
\vskip 0.75 true cm

\end{center}

{\leftskip=50pt \rightskip=50pt \noindent We show at  one-loop and first order in the noncommutativity parameters that in 
any noncommutative GUT inspired theory the total contribution to the fermionic four point functions coming only from the interaction between fermions and gauge bosons, though not UV finite by power counting, is UV finite at the end of the  
day. We also show that this is at odds with the general case for noncommutative gauge theories --chiral or otherwise-- defined by means of Seiberg-Witten maps that are the same --barring the gauge group representation-- for left-handed spinors as for right-handed spinors. We 
believe that the results presented in this paper tilt the scales to the side 
of noncommutative GUTS and noncommutative GUT inspired versions of the Standard Model. \par }

\vspace{20pt} \noindent
\end{titlepage}


\setcounter{page}{2}
\section{Introduction}

	The formulation of noncommutative gauge theories by means of Seiberg-Witten maps, also called the enveloping-algebra approach, allows to construct gauge theories for arbitrary groups and representations. This includes, in particular, noncommutative versions of the Standard Model \cite{Calmet:2001na}, Grand Unification Theories \cite{Aschieri:2002mc}, and theories compatible with the latter, meaning that they can be embedded in a GUT theory in a way consistent with its noncommutative and ordinary gauge symmetries. Since in GUT theories generically  the fermion fields are arranged in chiral representations of the gauge group that include both the left-handed components and the conjugate of the right-handed components of some of the Standard Model fermion fields, the GUT compatibility requirement amounts to demand that both left-handed fermions and the conjugates of right-handed fermions transform identically under noncommutative gauge transformations \cite{Aschieri:2002mc}. This implies that the left and right handed noncommutative fermion fields are related to their ordinary counterparts through different Seiberg-Witten maps -differing by a change of sign in the noncommutativity parameters. This rules out theories with maps defined in terms of Dirac fermions, among them the Standard Model of ref.~\cite{Calmet:2001na}; nevertheless, GUT-compatible versions of these models can be constructed \cite{Aschieri:2002mc}.
	
	It is known that noncommutative theories defined by means of Seiberg-Witten maps have anomaly cancellation conditions identical to their commutative counterparts  \cite{Brandt:2003fx},  which encourages the study of other quantum properties such as renormalisability for anomaly free theories, including the noncommutative versions of the Standard Model and GUT theories alluded to before. Most of the work so far has concerned theories including  noncommutative Dirac fermions, and thus not GUT-compatible. In general, it is known that the gauge sector of noncommutative theories is one-loop renormalisable at least at first order in the noncommutativity parameters, even when including the loop effects of scalars and Dirac fermions in arbitrary representations and  Majorana fermions in the adjoint \cite{Wulkenhaar:2001sq,Buric:2002gm,Buric:2004ms,Buric:2005xe,Latas:2007eu,Buric:2006wm,Martin:2006gw,Martin:2007wv,Martin:2009mu}. However, if the matter sector comprises Dirac fermions, the renormalisability is spoilt by divergences in the fermionic four point functions, as shown in the U(1) case and for SU(2) in the fundamental representation in refs.~\cite{Wulkenhaar:2001sq,Buric:2002gm,Buric:2004ms}. It is still a pending task to find a noncommutative gauge theory with a renormalisable one-loop effective action to first order in the noncommutativity parameters, involving matter in arbitrary representations. Supersymmetry on the side of noncommutative fields has been shown to be of help to yield one-loop renormalisable models, at least for U(N) and SU(N) ${\cal N}=1$ superYang-Mills \cite{Martin:2009mu}, but only involving gauginos in the adjoint representation. Regarding chiral fermions, it has been shown that the pathological four fermion divergences cancel for a single Weyl fermion field, both in the U(1) case and for SU(2) in the fundamental representation \cite{Buric:2007ix}. It is also worth noticing that the renormalisability of the matter sector of a noncommutative generalisation of the Standard Model in the enveloping-algebra approach has not been addressed so far. 

	 In this paper we thoroughly extend the seminal discoveries made, for U(1) and SU(2) in the fundamental representation, in ref.~\cite{Buric:2007ix}. We do so by considering arbitrary groups and representations and without restricting to only noncommutative left-handed 
fermions --as shown in ref. \cite{Aschieri:2002mc}, in noncommutative gauge theories defined by means of the Seiberg-Witten map, a theory with both left-handed  and right-handed multiplets  may not be equivalent to a theory formulated in terms of a single  
left-handed multiplet. We show that the divergences of the 4-point fermionic Green functions are absent at one-loop and first order in $\theta$, in GUT-compatible noncommutative gauge theories with an arbitrary gauge group,  in which the fermion fields belong to an anomaly-free --see ref. \cite{Brandt:2003fx}-- arbitrary representation.  We also show that, when the noncommutative theory is not GUT compatible, the divergences of the 4-point fermionic Green functions do not cancel.
	 
	  Our result clearly favours GUT-inspired noncommutative theories over those formulated in terms of noncommutative Dirac fermions, and stimulates the hope that one-loop renormalisable noncommutative gauge theories can still be constructed. In particular, since our calculation applies to the GUT inspired versions of the Standard Model of ref.~\cite{Aschieri:2002mc} when the Higgs interactions are neglected, our results represent the first albeit only partial study of the renormalisability of the matter sector of a theory directly related with a noncommutative generalisation of the Standard Model in the formalism that makes use of Seiberg-Witten  maps. The renormalisability of the full matter sector, with the inclusion of the Higgs interactions, is still an open problem.

\section{Theories and computations}

	We will study simultaneously theories that are GUT compatible and not GUT compatible. The action of these theories is given by
\begin{align}\label{S}
&{S_{\pm}}=\idx-\frac{1}{2g^2}\Tr F_{\m\n}\star F^{\m\n}+\Psib_L i\Dirac_{L} \Psi_L+\Psib_{R\pm} i\Dirac^{\pm}_{R} \Psi_{R\pm},\\
&\nonumber F_{\m\n}=\partial_\m A_\n-\partial_\n A_\m-i[A_\m, A_\n]_\star,\quad D_{L,\,\m}\psi_L=\partial_\m\Psi_L-i \rho_{L}(A_\m)\star\Psi_L,\\
\nonumber&D^+_{R,\m}\Psi_R=\partial_\m\Psi_R-i \rho_{R}(A_\m)\star\psi_R,\quad D^-_{R,\m}\Psi_R^{\top}=\partial_\m\Psi^{\top}_{R}+i\Psi^{\top}_{R}\star \rho_{R}^\star A_\m,
\end{align}
where the choice of ``$-$" (``$+$") corresponds to a theory which is (not) GUT-compatible; the $\pm$ theories differ in the Seiberg-Witten map for $\Psi_R$. The noncommutative product $\star$ is the usual Moyal product, 
\begin{align*}
a\star b=a \exp\Big[\frac{i}{2}\th^{\m\n}\overleftarrow{\partial}_\m\overrightarrow{\partial}_\n\Big]b,
\end{align*}
and $\rho_{L,R}$ designates the representations of the left and right fermionic fields $\Psi_L,\,\Psi_{R\pm}$, $\rho^\star_R$ being the conjugate representation of $\rho_R$.  The noncommutative fields $A_\m$ and $\Psi_L,\,\Psi_{R\pm}$ are defined in terms of their  ordinary counterparts $a_\m,\,\psi_L,\,\psi_R$ through the following   Seiberg-Witten maps
\begin{align}\nonumber
A_\mu&=a_\m+\frac{1}{4}\th^{\a\b}\{\partial_\a \am+f_{\a\m},\ab\}+O(\th^2),\\
\label{SW}\Psi_L&=\psi_L-\frac{1}{2}\th^{\a\b}\rho_L(a_\a)\Db\psi_L+\frac{i}{4}\th^{\a\b}\rho_L(a_\a) \rho_L(a_\b)\psi_L+O(\th^2),\\
\nonumber \Psi_{R\pm}&=\psi_R\mp\frac{1}{2}\th^{\a\b}\rho_R(a_\a)\Db\psi_R\pm\frac{i}{4}\th^{\a\b}\rho_R(a_\a) \rho_R(a_\b)\psi_R+O(\th^2).
\end{align}
Note that the GUT-compatible maps of $\Psi_L$ and $\Psi_{R-}$ differ by a change of sign in $\theta^{\m\n}$. This has as a consequence that the conjugate $\Psi_{R-}^C$ of $\Psi_{R-}$ has the same Seiberg-Witten map in terms of the left-handed $\psi_R^C$ as the map for $\Psi_L$ in eq.~\eqref{SW}, with $\rho_L$ substituted by  $\rho_R$ (note that $\rho_R^\star(a_\mu)=-\rho_R(a_\mu)$)  \cite{Aschieri:2002mc}. This is equivalent to saying that both $\Psi_L$ and $\Psi_{R-}^C$, barring the representation, transform identically under noncommutative gauge transformations and thus can be embedded in a single multiplet, as required in GUT theories. On the other hand, in the case which is not GUT-compatible, the noncommutative Weyl fermions $\Psi_L$ and $\Psi_{R+}$ transform in the same way under noncommutative gauge transformations --though they might do so  under different representations-- and can thus be embedded in a noncommutative Dirac fermion.  It should be  pointed out that the trace operation $\frac{1}{g^2}\Tr$ in eq.~\eqref{S} is in general ambiguous and different choices can lead to inequivalent models; the results of this paper do not depend on the choice of trace and therefore we will leave it unspecified.

Expanding the two actions $S_\pm$ in eq.~\eqref{S} with the Seiberg-Witten maps of eq.~\eqref{SW}, the two theories that are obtained are inequivalent. Their right-handed fermionic sectors differ, and it can be shown that the following relationship holds \cite{Aschieri:2002mc}
\begin{align*}
\idx\Psib_{R-} i\Dirac^{-}_{R} \Psi_{R-}=\left.\idx\Psib_{R+} i\Dirac^{+}_{R} \Psi_{R+}\right|_{\th\rightarrow-\th}.
\end{align*}

We consider the ordinary field $a_\m$ taking values in the Lie algebra of an arbitrary semisimple group of the form $G_1\times\dots\times G_N$ with $G_i$ simple for $i=1\dots s$ and abelian for $i=s+1,\dots,N$:
\begin{equation}\label{aexp}
a_\m=\sum_{k=1}^sg_k(a^k_\m)^a(T^k)^a+\sum_{l=s+1}^Ng_l a_\m^lT^l\equiv\sum_{m=1}^N\sum_{(a)}g_m (a_\m)^{(a)}(T^m)^{(a)},
\end{equation}
where the $T$'s are generators of unitary representations of the group factors.  $\rho_{L,R}$ can be taken as arbitrary, anomaly free unitary representations, which might be expressed as a direct sum of irreducible representations, $\rho_{L,R}=\bigoplus_{r=1}^F \rho^{r}_{L,R}$.  Accordingly, the fermion fields can be expressed as a direct sum of irreducible multiplets, $\Psi_{L,R}=\bigoplus_{r=1}^F \Psi_{L,R}^{r}, \psi_{L,R}=\bigoplus_{r=1}^F \psi_{L,R}^{r}$. Each $\psi_{L,R}^{r}$ in an irreducible representation carries multi-indices  $I=i_1\dots i_s$ for the different group factors. In multi-index notation we can define the generators as follows
\begin{equation*}
\begin{array}{l}
\rho(T^i)=\bigoplus_r\rho^r(T^i),\quad i=1,\dots,N,\\
\rho^r((T^k)^a)_{IJ}=\delta_{i_1 j_1}\cdots\rho^r((T^k)^a)_{i_k j_k}\cdots \delta_{i_s j_s},\quad k=1,\dots, s,\\
\rho^r(T^l_{IJ})= \delta_{i_1 j_1}\cdots \delta_{i_s j_s}\rho^r(Y^l),\quad l=s+1,\dots,N.
\end{array}
\end{equation*}
 Note that, in the GUT-compatible case, since $\Psi_{R-}^C$ behaves under noncommutative gauge transformations as $\Psi_L$, both can be combined into a single reducible left-handed multiplet, and one could study the theory by considering just a left-handed fermion field. However, this would not allow for a direct comparison with the non-GUT-compatible case.

 Our model  can  accommodate the GUT-inspired theories of ref.~\cite{Aschieri:2002mc}. For example, the noncommutative versions of the SU(5) or SO(10) GUTS can be obtained by dropping $\Psi_R$ and considering a  representation for $\Psi_L$ which includes the left-handed Standard Model fields and the conjugates of the right-handed ones. One can also obtain
 the GUT inspired noncommutative $\rm QED_-$ and the SO(10)-embeddable Standard Model of ref.~\cite{Aschieri:2002mc}, either by considering both $\Psi_L$ and $\Psi_R$ fields or  a single left-handed field.

	We quantise the ordinary fields $a_\m,\psi_{L,R}$ of the theory by defining the functional generator in terms of an expansion in Feynman diagrams. To simplify the computations so as to be able to formulate the Feynman rules in terms of ordinary Dirac fermions --with $\psi_{L}$ and $\psi_R$ embedded in different Dirac fermions  $\psi$ and $\psi'$-- we add  non-interacting additional right and left handed fermion multiplets.
\begin{equation}
S\rightarrow S'= S+\idx\bar{\tilde\psi}_L i\prslash\tilde\psi_L+\idx\bar{\tilde\psi}_R i\prslash\tilde\psi_R,	\quad \psi=\left[\begin{array}{l}\tilde\psi_R\\
\psi_L\end{array}\right],\psi'=\left[\begin{array}{l}\psi_R\\
\tilde\psi_L\end{array}\right].\quad\label{Sshift}
\end{equation}
	This clearly does not affect the interactions of the original fermions. The original action of eq.~\eqref{S}, after expanding it to order $\th$ with the Seiberg-Witten maps of eq.~\eqref{SW}, is of the form
\begin{equation*}
S_{\rm fermion}=\idx \bar\psi_L( i\prslash+\gamma^\m{\cal O}_\m[a,\partial,\th])\psi_L+\psi_R( i\prslash+\gamma^\m{\cal O}_{\pm\m}[a,\partial,\th])\psi_R;
\end{equation*}
then, after the modification of eq.~\eqref{Sshift} we have
\begin{equation*}
S'_{\rm fermion}=\idx \bar\psi( i\prslash+\gamma^\m P_L{\cal O}_\m[a,\partial,\th])\psi+\bar\psi'( i\prslash+\gamma^\m P_R{\cal O}_{\pm\m}[a,\partial,\th])\psi'.
\end{equation*}
Since we shall
use dimensional regularisation --with $D=4+2\epsilon$-- and there is a
$\gamma_5$ in the interaction vertices, one has to specify what one means by 
$\gamma_5$ in dimensional regularisation and thus, in turn, one has to state the dimensional regularisation scheme chosen to regularise the theory. Here, we shall use the scheme for defining $\gamma_5$ in the 
dimensionally regularised theory put forward  by 't Hooft and Veltman \cite{'tHooft:1972fi} and formulated 
rigurously by Breitenlohner and Maison \cite{Breitenlohner:1977hr}: the famous BMHV scheme. In this
scheme the dimensionally regularised action --and with it, the Feynman rules-- are not  
unique. There is an infinity of  dimensionally regularised actions which differ
from one another by evanescent operators \cite{Martin:1999cc}. Here, we shall  
follow ref. \cite{Martin:1999cc} and keep all the vector indices in interaction 
vertices ``four-dimensional'', i.e., contracted with the ``barred'' metric $\bar{g}_{\mu\nu}$. In keeping with the BHMV scheme, we shall also define the dimensionally regularised $\theta^{\mu\nu}$ as being ``four-dimensional'' --see ref. \cite{Martin:2005gt}, for details. Of course, the dimensionally regularised free propagators are the canonical ``D-dimensional'' ones \cite{Breitenlohner:1977hr}.

\begin{figure}[h]
\begin{minipage}{0.3\textwidth}
\psfrag{p}{$p$}\psfrag{q}{$q$}\psfrag{k}{$k$}
\psfrag{i s}{$m s$}\psfrag{j t}{$n t$}\psfrag{k,a,m}{$i,(a),\mu$}
\includegraphics[scale=0.5]{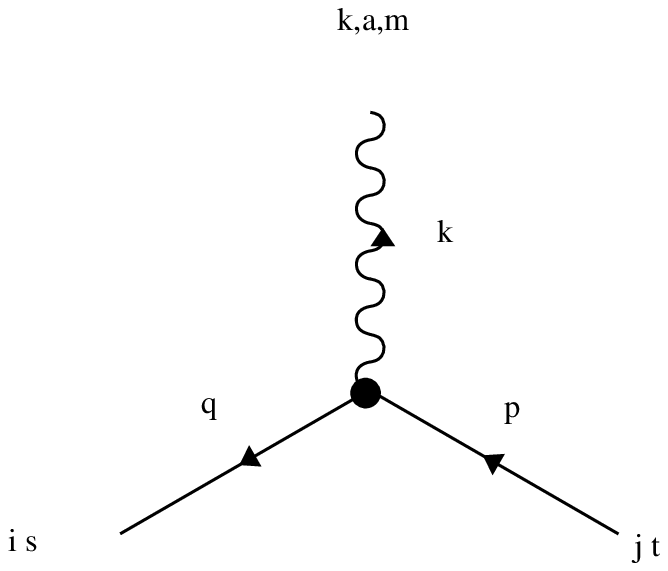}%
\end{minipage}%
\begin{minipage}{0.68\textwidth}%
\begin{align*}
\leftrightarrow\frac{1}{2}(\bg^\n P_L)_{mn}\th^{\a\b}g_i\rho_L
(T^i)^{(a)}_{st}[-(\bq_\n \bp_\b-\bq_\b \bp_\n){\bdelta^\m}_\a)-\bk_\a \bp_\b{\bdelta^\m}_\n]
\end{align*}
\end{minipage}
\vskip0.2cm
\begin{minipage}{0.27\textwidth}
\psfrag{p}{$p$}\psfrag{q}{$q$}\psfrag{k1}{$k_1$}\psfrag{k2}{$k_2$}
\psfrag{i s}{$m s$}\psfrag{j t}{$n t$}\psfrag{k,a,m}{$i,(a),\mu$}
\psfrag{m1}{$i,(a),\mu$}\psfrag{m2}{$j,(b),\nu$}
\includegraphics[scale=0.7]{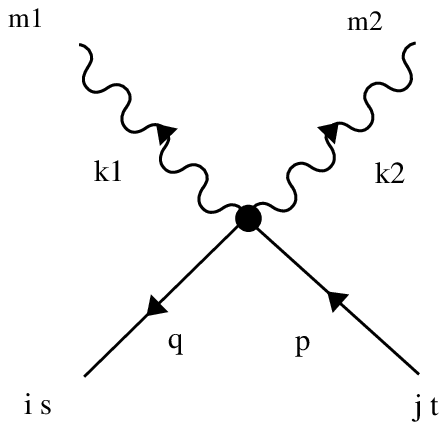}%
\end{minipage}%
\begin{minipage}{0.71\textwidth}%
\begin{align*}
\leftrightarrow&-\frac{1}{4}(\bg^\s P_L)_{mn}\th^{\a\b}g_ig_j[\{\rho_L(T^i)^{(a)},\rho_L(T^j)^{(b)}\}_{st}(-(\bk_1-\bk_2)_\s\bdelta^{\m}_\a\bdelta^\n_\b\\
&+2\bk_{1\a}\bdelta^\m_\s\bdelta^\n_\b+2\bk_{2\a}\bdelta^\n_\s\bdelta^\m_\b-(\bq-\bp)_\b(\bdelta^\m_\a\bdelta^\n_\s+\bdelta^\n_\a\bdelta^\m_\s))\\
&+[\rho_L(T^i)^{(a)},\rho_L(T^j)^{(b)}]_{st}((\bq+\bp)_\s\bdelta^\m_\a\bdelta^\n_\b-(\bq+\bp)_\b(\bdelta^\m_\a\bdelta^\n_\s-\bdelta^\n_\a\bdelta^\m_\s))].
\end{align*}
\end{minipage}
\caption{Feynman rules of the noncommutative interactions relevant to our calculations, involving the Dirac fermion $\psi$. }
\label{f:1}
\end{figure}

\begin{figure}[h]
\begin{minipage}{0.3\textwidth}
\psfrag{p}{$p$}\psfrag{q}{$q$}\psfrag{k}{$k$}
\psfrag{i s}{$m s$}\psfrag{j t}{$n t$}\psfrag{k,a,m}{$i,(a),\mu$}\psfrag{pm}{$\tiny \pm$}
\includegraphics[scale=0.5]{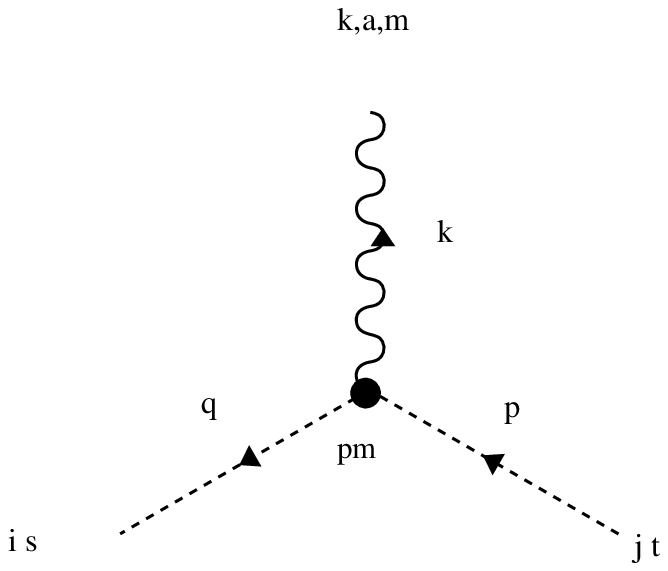}%
\end{minipage}%
\begin{minipage}{0.68\textwidth}%
\begin{align*}
\leftrightarrow\pm\frac{1}{2}(\bg^\n P_R)_{mn}\th^{\a\b}g_i\rho_R(T^i)^{(a)}_{st}[-(\bq_\n \bp_\b-\bq_\b \bp_\n)
{\bdelta^\m}_\a)-\bk_\a \bp_\b{\bdelta^\m}_\n]
\end{align*}
\end{minipage}
\vskip0.2cm
\begin{minipage}{0.27\textwidth}
\psfrag{p}{$p$}\psfrag{q}{$q$}\psfrag{k1}{$k_1$}\psfrag{k2}{$k_2$}
\psfrag{i s}{$m s$}\psfrag{j t}{$n t$}\psfrag{k,a,m}{$i,(a),\mu$}
\psfrag{m1}{$i,(a),\mu$}\psfrag{m2}{$j,(b),\nu$}\psfrag{pm}{$\pm$}
\includegraphics[scale=0.7]{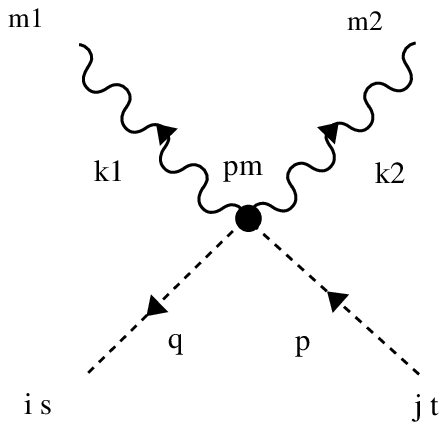}%
\end{minipage}%
\begin{minipage}{0.71\textwidth}%
\begin{align*}
\leftrightarrow&\mp\frac{1}{4}(\bg^\s P_R)_{mn}\th^{\a\b}g_ig_j[\{\rho_R(T^i)^{(a)},\rho_R(T^j)^{(b)}\}_{st}
(-(\bk_1-\bk_2)_\s\bdelta^{\m}_\a\bdelta^\n_\b\\
&+2\bk_{1\a}\bdelta^\m_\s\bdelta^\n_\b+2\bk_{2\a}\bdelta^\n_\s\bdelta^\m_\b-(\bq-\bp)_\b
(\bdelta^\m_\a\bdelta^\n_\s+\bdelta^\n_\a\bdelta^\m_\s))\\
&+[\rho_R(T^i)^{(a)},\rho_R(T^j)^{(b)}]_{st}((\bq+\bp)_\s\bdelta^\m_\a\bdelta^\n_\b-(\bq+\bp)_\b
(\bdelta^\m_\a\bdelta^\n_\s-\bdelta^\n_\a\bdelta^\m_\s))].
\end{align*}
\end{minipage}
\caption{Feynman rules of the noncommutative interactions relevant to our calculations, involving the Dirac fermion $\psi'$. }
\label{f:2}
\end{figure}

\begin{figure}[h]\centering
\psfrag{p}{$p$}\psfrag{q}{$q$}\psfrag{r}{$r$}\psfrag{s}{$s$}
\psfrag{i}{$i s$}\psfrag{j}{$j t$}\psfrag{k}{$k u$}
\psfrag{l}{$l v$}\psfrag{A1}{$A_1$}\psfrag{A2}{$A_2$}\psfrag{A3}{$A_3$}\psfrag{A4}{$A_4$}\psfrag{per2}{+3 perm.}
\psfrag{B}{$B_k,k=1...16$}\psfrag{C}{$C_k,k=1...16$}
\psfrag{Ap1}{$A'_1$}\psfrag{Ap2}{$A'_2$}\psfrag{Ap3}{$A'_3$}\psfrag{Ap4}{$A'_4$}\psfrag{per2}{+3 perm.}
\psfrag{Bp}{$B'_k,k=1...16$}\psfrag{Bpp}{$B''_k,k=1...8$}
\psfrag{App1}{$A''_1$}\psfrag{App2}{$A''_2$}\psfrag{pm}{$\pm$}
\includegraphics[scale=1]{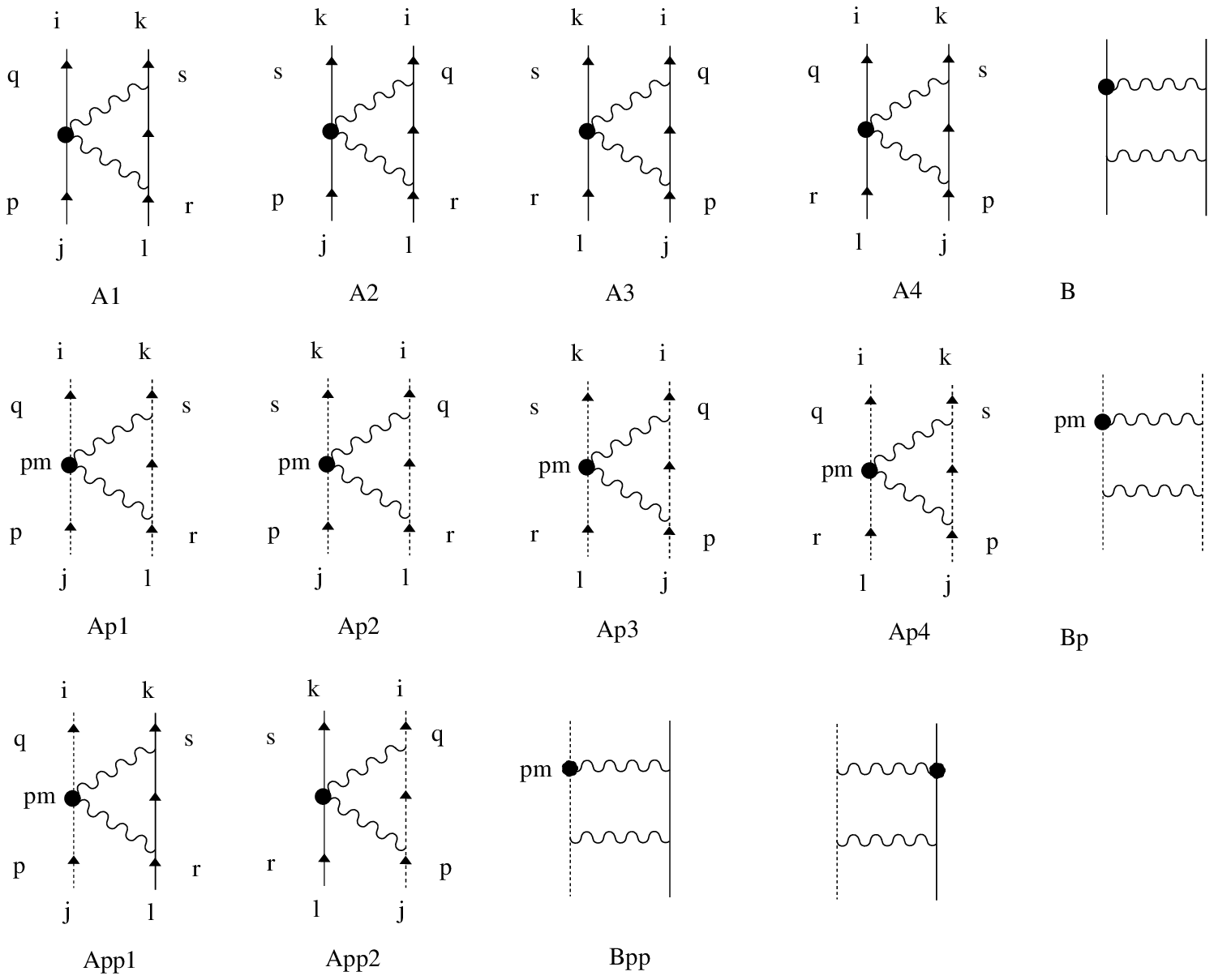}%
\caption{Diagrams contributing to  the fermionic four point functions at order $\th$.}
\label{f:3}
\end{figure}

 The rules involving noncommutative parameters that are relevant for the calculation of one-loop four point diagrams involving fermions at order $\th$ are shown in Figs.~\ref{f:1} and \ref{f:2}. The diagrams that contribute are displayed in Fig.~\ref{f:3}.

Let us explain the notation of the Feynman rules. First, solid lines represent propagators of the Dirac fermion $\psi$, while dashed lines represent propagators of $\psi'$. The indices $i,(a),\m$ on each gauge field leg denote, respectively, that the gauge field belongs to the $i$'th semisimple subalgebra in the expansion of eq.~\eqref{aexp}, and, in case the subalgebra is non-abelian, it is the component along the $a$'th generator; finally,  $\m$ is the Lorentz index. In the Feynman rules, the indices $i$ and $j$ are not summed over.  In a fermionic leg with labels``$nt$", ``$n$"  refers to the Dirac index, while ``$t$" is the index of the full representation $\rho=\bigoplus\rho^r$. Finally, ``barred'' ($\bg_\mu$, $\bp_\mu$, etc.) objects are defined as in refs. \cite{Breitenlohner:1977hr, Martin:1999cc}.

From the Feynman rules of Figs.~\ref{f:1} and \ref{f:2}, it is clear that the noncommutative 3-point vertices of the box diagrams in Fig.~\ref{f:3} always involve one external momentum; it is easily seen that this makes them finite by power counting, and thus
the diagrams $B$ and $C$ have zero divergences,
\begin{align*}
B_k^{\rm div}={B'_k}^{\rm div}={B''_k}^{\rm div}=0.
\end{align*}
On the other hand, after some Dirac algebra the results for $A_i$ can be expressed as follows
\begin{align*}
A_1^{\rm div}=&\frac{-3i}{128\pi^2\epsilon}\!\!\!\sum_{m,n,(a),(b)}\!\!\!\!\!g_m^2g_n^2\{\rho_L(T^m)^{(a)},\rho_L(T^n)^{(b)}\}_{st}\{\rho_L(T^m)^{(a)},\rho_L(T^n)^{(b)}\}_{uv}\tilde\th_{\m\n}(\bg^\m P_L)_{ij}(\bg^\n P_L)_{kl},\\
A_2^{\rm div}=&\frac{-3i}{128\pi^2\epsilon}\!\!\!\sum_{m,n,(a),(b)}\!\!\!\!\!g_m^2g_n^2\{\rho_L(T^m)^{(a)},\rho_L(T^n)^{(b)}\}_{ut}\{\rho_L(T^m)^{(a)},\rho_L(T^n)^{(b)}\}_{sv}\tilde\th_{\m\n}(\bg^\m P_L)_{kj}(\bg^\n P_L)_{il},\\
A_3^{\rm div}=&\frac{-3i}{128\pi^2\epsilon}\!\!\!\sum_{m,n,(a),(b)}\!\!\!\!\!g_m^2g_n^2\{\rho_L(T^m)^{(a)},\rho_L(T^n)^{(b)}\}_{uv}\{\rho_L(T^m)^{(a)},\rho_L(T^n)^{(b)}\}_{st}\tilde\th_{\m\n}(\bg^\m P_L)_{kl}(\bg^\n P_L)_{ij},\\
A_4^{\rm div}=&\frac{-3i}{128\pi^2\epsilon}\!\!\!\sum_{m,n,(a),(b)}\!\!\!\!\!g_m^2g_n^2\{\rho_L(T^m)^{(a)},\rho_L(T^n)^{(b)}\}_{sv}\{\rho_L(T^m)^{(a)},\rho_L(T^n)^{(b)}\}_{ut}\tilde\th_{\m\n}(\bg^\m P_L)_{il}(\bg^\n P_L)_{kj},
\end{align*}
where $\tilde\th^{\m\n}\equiv\frac{1}{2}\epsilon^{\m\n\a\b}\th_{\a\b}$ and the summations over the Lie Algebra indices $m,(a)$ have been explicitly indicated -see eq.~\eqref{aexp}- while repeated Lorentz indices imply a sum as usual. It is clear that
\begin{align*}
A_1^{\rm div}+A_3^{\rm div}=0,\quad A_2^{\rm div}+A_4^{\rm div}=0,
\end{align*}
so that the total sum cancels. Similarly, for the diagrams $A'_i$ we have that
\begin{align*}
{A'_k}^{\rm div,\pm}=\mp A_k^{\rm div}|_{L\rightarrow R},
\end{align*}
and they also add up to zero. Finally, the diagrams $A''_k$ have the following divergent contributions
\begin{align*}
{A''_1}^{\rm\pm div}\!\!\!=&\frac{\mp 3i}{128\pi^2\epsilon}\!\!\!\!\sum_{m,n,(a),(b)}\!\!\!\!\!g_m^2g_n^2\{\rho_R(T^m)^{(a)},\rho_R(T^n)^{(b)}\}_{st}\{\rho_L(T^m)^{(a)},\rho_L(T^n)^{(b)}\}_{uv}\tilde\th_{\m\n}(\bg^\m P_R)_{ij}(\bg^\n P_L)_{kl},\\
{A''_2}^{\rm \pm div}\!\!\!=&\frac{-3i}{128\pi^2\epsilon}\!\!\!\!\sum_{m,n,(a),(b)}\!\!\!\!\!g_m^2g_n^2\{\rho_R(T^m)^{(a)},\rho_R(T^n)^{(b)}\}_{st}\{\rho_L(T^m)^{(a)},\rho_L(T^n)^{(b)}\}_{uv}\tilde\th_{\m\n}(\bg^\m P_R)_{ij}(\bg^\n P_L)_{kl}.
\end{align*}
In the $S_+$ theory, which is not GUT-compatible, the two previous divergent contributions are equal and do not cancel. It should be noted that this happens even for chiral theories, as long as both left-handed and right-handed fields are present in the original action. However, in the GUT-compatible $S_-$ theory the divergences cancel each other. This fits with the cancellation of the divergences of the $A_i,A'_i$ diagrams, which only involve interactions of a single ordinary fermion field: as was said before, in the GUT-compatible case the noncommutative fields $\Psi_L$ and $\Psi_R^C$ can be embedded in a single fermion field, whose ordinary counterpart can be completed into a Dirac field by adding a spectator fermion. Thus the results in the GUT-compatible case should be equivalent to those when only one of the Dirac fermions $\psi$ or $\psi'$ is present.

\section{Conclusions}
The main conclusion of the computations carried out in this paper is that if we tailor our noncommutative gauge theories according to
the GUT framework, the UV divergent behaviour of the theory is improved with regard to that of the corresponding non-GUT-compatible  construction.
We believe that theories such as the GUT-inspired noncommutative SM and the SO(10) GUT theory of  ref. \cite{Aschieri:2002mc} deserve further
analysis with regard to either their complete one-loop renormalisability or their phenomenological consequences, which may be tested at the LHC. A particularly interesting -though more involved from the computational standpoint-- and completely open issue is the UV properties the GUT-inspired theories
when a Higgs sector and Yukawa couplings are included. Indeed, Higgs interactions are introduced by means of a hybrid Seiberg-Witten map \cite{Calmet:2001na, Aschieri:2002mc} and, as it happens  for the interactions studied in this paper, power-counting
does not  forbid that  these new interactions give rise  to new UV divergent 4-point fermionic contributions.

\section{Acknowledgements}

This work has been financially supported in part by MICINN under grant
FPA2008-04906, and by the National Science Foundation under Grant No. PHY05-51164. The work of C.~Tamarit has also received financial support from MICINN and the Fulbright Program through grant 2008-0800.


\end{document}